\documentstyle[12pt,epsfig]{article}

\def\dtpAv{  .007~}
\def\dtnAv{  .004~}

\def\dtpAvEr{  .004~}
\def\dtnAvEr{  .010~}

\def\ChiGwwd{0.9~}
\def\ChiGwwp{1.3~}
\def\ChiZerod{0.9~}
\def\ChiZerop{1.5~}
\def\dtp{ 0.005~}
\def\dtpEr{ 0.008~}
\def\dtd{ 0.008~}
\def\dtdEr{ 0.005~}
\def\BCp{-0.022~}
\def\BCpEr{ 0.071~}
\def\BCd{ 0.023~}
\def\BCdEr{ 0.044~}
\def\BCAvp{-0.015~}
\def\BCAvpEr{ 0.026~}
\def\BCAvd{ 0.010~}
\def\BCAvdEr{ 0.039~}
\def\ELT{-0.015~}
\def\ELTEr{ 0.036~}
\def\ELTAv{ 0.003~}
\def\ELTAvEr{ 0.022~}

\begin{document}
\thispagestyle{empty}
\renewcommand{\thefootnote}{\fnsymbol{footnote}}
\begin{flushright}
{\small
SLAC-PUB-7983\\
Jan. 1999 \\
}
\end{flushright}
\vspace{.8cm}
\begin{center}{\bf\large

Measurement of the Proton and Deuteron Spin Structure Functions 
g$_{2}$ and Asymmetry A$_{2}$\footnote{Work supported by
Department of Energy contract  DE--AC03--76SF00515.}}
\break
\vskip 1cm
{The E155 Collaboration \break  
P.~L.~Anthony,$^{16}$  
R.~G.~Arnold,${^1}$
T.~Averett,$^{5,\dag\dag}$
H.~R.~Band,$^{21}$
M.~C.~Berisso,$^{12}$
H.~Borel,$^7$
P.~E.~Bosted,${^1}$
S.~L.~B${\ddot {\rm u}}$ltmann,$^{19}$
M.~Buenerd,$^{16,\dag}$
T.~Chupp,$^{13}$
S.~Churchwell,$^{12,\ddag}$
G.~R.~Court,$^{10}$
D.~Crabb,$^{19}$
D.~Day,$^{19}$
P.~Decowski,$^{15}$
P.~DePietro,$^1$
R.~Erbacher,$^{16,17}$
R.~Erickson,$^{16}$
A.~Feltham,$^{19}$
H.~Fonvieille,$^3$
E.~Frlez,$^{19}$
R.~Gearhart,$^{16}$
V.~Ghazikhanian,$^6$  
J.~Gomez,$^{18}$
K.~A.~Griffioen,$^{20}$
C.~Harris,$^{19}$
M.~A. Houlden,$^{10}$
E.~W.~Hughes,$^5$
C.~E.~Hyde-Wright,$^{14}$
G.~Igo,$^6$
S.~Incerti,$^3$
J.~Jensen,$^5$
J.~K.~Johnson,$^{21}$
P.~M.~King,$^{20}$
Yu.~G.~Kolomensky,$^{5,12}$
S.~E.~Kuhn,$^{14}$
R.~Lindgren,$^{19}$
R.~M.~Lombard-Nelsen,$^7$
J.~Marroncle,$^7$
J.~McCarthy,$^{19}$
P.~McKee,$^{19}$
W.~Meyer,$^{4}$
G.~S.~Mitchell,$^{21}$
J.~Mitchell,$^{18}$
M.~Olson,$^{9,\S\S}$
S.~Penttila,$^{11}$
G.~A.~Peterson,$^{12}$
G.~G.~Petratos,$^9$ 
R.~Pitthan,$^{16}$
D.~Pocanic,$^{19}$
R.~Prepost,$^{21}$   
C.~Prescott,$^{16}$
L.~M.~Qin,$^{14}$
B.~A.~Raue,$^{8}$
D.~Reyna,$^{1,\heartsuit}$
L.~S.~Rochester,$^{16}$
S.~Rock,$^1$
O.~Rondon-Aramayo,$^{19}$
F.~Sabatie,$^7$
I.~Sick,$^2$
T.~Smith,$^{13}$
L.~Sorrell,$^1$
F.~Staley,$^7$
S.~St.Lorant,$^{16}$
L.~M.~Stuart,$^{16,\S}$
Z.~Szalata,$^1$
Y.~Terrien,$^7$
A.~Tobias,$^{19}$
L.~Todor,$^{14}$ 
T.~Toole,$^1$
S.~Trentalange,$^{6}$
D.~Walz,$^{16}$
R.~C.~Welsh,$^{13}$
F.~R.~Wesselmann,$^{14}$
T.~R.~Wright,$^{21}$
C.~C.~Young,$^{16}$
M.~Zeier,$^2$
H.~Zhu,$^{19}$
B.~Zihlmann$^{19}$  
}
\vskip .5cm
{
{$^{1}$American University, Washington, D.C. 20016}  \break
{$^{2}$Institut f${\ddot u}$r Physik der Universit${\ddot a}$t Basel, CH-40=
56 Basel, Switzerland} \break
{$^{3}$University Blaise Pascal, LPC IN2P3/CNRS F-63170 Aubiere Cedex, Fran=
ce}
\break
{$^{4}$Ruhr-Universit${\ddot a}$t Bochum, Universit${\ddot a}$tstr. 150,=20
Bochum, Germany} \break
{$^{5}$California Institute of Technology, Pasadena, California 91125}
\break
{$^{6}$University of California, Los Angeles, California 90095}
\break
{$^{7}$DAPNIA-Service de Physique Nucleaire, CEA-Saclay,
F-91191 Gif/Yvette Cedex, France} \break
{$^{8}$Florida International University, Miami, Florida 33199.} \break
{$^{9}$Kent State University, Kent, Ohio 44242} \break
{$^{10}$University of Liverpool, Liverpool L69 3BX, United Kingdom } \break
{$^{11}$Los Alamos National Laboratory, Los Alamos, New Mexico 87545} \break
{$^{12}$University of Massachusetts, Amherst, Massachusetts 01003} \break
{$^{13}$University of Michigan, Ann Arbor, Michigan 48109} \break
{$^{14}$Old Dominion University, Norfolk, Virginia 23529} \break
{$^{15}$Smith College, Northampton, Massachusetts 01063} \break
{$^{16}$Stanford Linear Accelerator Center, Stanford, California 94309 } \break
{$^{17}$Stanford University, Stanford, California 94305} \break
{$^{18}$Thomas Jefferson National Accelerator Facility, Newport News, Virginia
23606} \break
{$^{19}$University of Virginia, Charlottesville, Virginia 22901} \break
{$^{20}$The College of William and Mary , Williamsburg, Virginia 23187} \break
{$^{21}$University of Wisconsin, Madison, Wisconsin 53706} \break
}

\end{center}
\vfill

\begin{center}
{\bf\large   
Abstract }
\end{center}
 
\begin{quote}
We have measured the spin structure functions g$_{2}^{p}$ and g$_{2}^{d}$  and 
the virtual photon asymmetries A$_{2}^{p}$ and A$_{2}^{d}$ over the 
kinematic range
$0.02\leq x \leq 0.8$ and $1.0 \leq Q^{2} \leq 30$ (GeV/c)$^{2}$  
by scattering 38.8
GeV longitudinally polarized electrons from transversely polarized NH$_3$ and 
$^6{\rm LiD}$ targets.
The absolute value of  A$_{2}$ is significantly smaller than
the $\sqrt{R}$ positivity limit over  the measured range, while
g$_2$ is  consistent with the twist-2 
Wandzura-Wilczek calculation.
We obtain results 
for the twist-3 reduced matrix elements  $d_{2}^{p}$,  $d_{2}^{d}$ and 
$d_{2}^{n}.$  The Burkhardt-Cottingham sum rule 
integral $\int$g$_{2}(x)dx$ is reported for the range 
$0.02\leq x \leq 0.8$.  
\end{quote}
\vfill
\begin{center} 
{\it To be submitted to Physics Letters} \break
\end{center}

%PACS numbers 13.60.Hb, 13.88.+e, 24.70.+s, 25.30.Fj
%Keywords: nucleon, spin structure, deep inelastic scattering, 
%          neutron, spin, structure function, transverse

\newpage 
\pagestyle{plain}

The deep inelastic spin structure functions of the nucleons, $g_{1}$ and 
$g_{2}$, depend on the spin distribution of the partons and their
correlations.  The function $g_1$ can be primarily understood in
terms of the quark parton model (QPM) and perturbative QCD 
with higher twist terms at low $Q^2.$   There
exists no such  picture for $g_2$. Feynman\cite{Feynman}
claimed that the transverse structure function $g_T =g_1 +g_2$  
had a simple parton interpretation in terms of the transverse polarization
of the quark spins  which is proportional to  quark masses.
However, $g_2$ is sensitive to higher twist effects such as 
quark-gluon correlations\cite{Vain}  and is not easily
interpreted in pQCD where such effects are not included.
By interpreting $g_{2}$ using the operator product expansion (OPE)
~\cite{Vain,Jaffe}, it is possible to study contributions to
the nucleon spin structure beyond the simple QPM.  The virtual
photon-nucleon asymmetry $A_2$ is proportional to $g_T/F_2$ where $F_2$ is
the unpolarized structure function.

The structure function $g_2$ can be written \cite{CPR}:
\begin{equation}
g_2(x,Q^2) = g_2^{WW}(x,Q^2) +\overline{g_2}(x,Q^2)
\end{equation}
where
\begin{eqnarray}
{\rm{g}}_{2}^{WW}(x,Q^{2})&=&-{\rm{g}}_{1}(x,Q^{2}) + \int_{x}^{1}
\frac{{\rm{g}}_{1}(y,Q^{2})}{y} dy.\nonumber \\ 
\overline{g_2}(x,Q^2) &=& -\int_x^1{\partial\over \partial y}\biggl({m\over M}h_T(y,Q^2)+\xi(y,Q^2)\biggr){dy\over y}  \nonumber
\label{eq:g2ww}
\end{eqnarray}
where  $x$ is the  Bjorken scaling variable and $Q^{2}$ is the absolute value
of the virtual photon four-momentum squared.
The twist-2 term $g_2^{WW}$ was derived by  Wandzura and Wilczek~\cite{g2ww}
and depends only on the well-measured 
$g_1$ \cite{SMC,E143,E142_2,E154,E155,Hermes}.
The function $h_T(x,Q^2)$ is an additional  twist-2
contribution ~\cite{Song,CPR} 
that depends on the transverse polarization density in the 
nucleon. The $h_T$ contribution to  $\overline{g_2}$,  is  suppressed 
by the ratio of the quark-to-nucleon mass~\cite{Song} 
and is thus small for up and down quarks 
and will be neglected in this analysis \cite{RR}. 
The twist-3 part, $\xi,$ comes from quark-gluon correlations and is the main
focus of our study.

The OPE allows us to write the hadronic matrix element in deep 
inelastic scattering in terms of a series of renormalized operators of
increasing twist~\cite{Vain,Jaffe}.  
The moments of $g_{1}$ and $g_{2}$ at fixed $Q^{2}$ can be related 
to the twist-2 and twist-3
reduced matrix elements, $a_n$ and $d_n$, and higher twist
terms which are suppressed by powers of $1/Q$. Neglecting quark mass terms:
\begin{eqnarray}
\label{eq:moments}
\int_0^1x^n{\rm{g}}_{1}(x,Q^{2})dx=\frac{a_{n}}{2} +
{\it O}(M^2/Q^2),
\;\;n=0,2,4,... \nonumber \\ 
\int_0^1x^n{\rm{g}}_2(x,Q^2)dx=\frac{n}{n+1}\frac{(d_n-a_n)}{2}
+ {\it O}(M^2/Q^2),
\;\;n=2,4,...
\end{eqnarray}
In these integrals
the contribution of $d_n$ is not suppressed relative to the twist-2 
contribution and thus can be easily extracted.
Neglecting $(1/Q)$ terms, the $d_n$ matrix elements can be written  as:
\begin{equation}
d_n= \frac{ 2(n+1)}{n}\int_0^1 x^n\overline{g_2}(x,Q^2)dx 
\label{eq:g2bar}
\end{equation}
and thus measures deviations of $g_2$ from the twist-2 $g_2^{WW}$ term.

The Burkhardt-Cottingham sum rule\cite{buco} for $g_2$ at large $Q^2$, 
\begin{eqnarray}
\int_0^1 g_2(x)dx=0\ ,\label{eq:bc}
\end{eqnarray}
was derived from virtual Compton scattering dispersion relations. It
does not follow from the OPE since the $n=0$ sum rule is not defined for 
$g_2.$ Its validity depends on the lack of singularities for $g_2$ at $x=0.$
The  Efremov-Leader-Teryaev (ELT) sum rule\cite{ELT} involves the valence 
quark contributions to $g_1$ and $g_2$:
\begin{eqnarray}
\int_0^1 x[g_1^V(x) + 2g_2^V(x)]dx=0\ .\label{eq:elt}
\end{eqnarray}
Assuming that the sea quarks are the same in protons and neutrons the sum
rule takes a form
$\int_0^1 x[g_1^p(x) + 2g_2^p(x) - g_1^n(x) - 2g_2^n(x) ]dx=0$ 
that we can apply to our data.

Measurements of g$_2$ and A$_2$ exist for the
proton and  deuteron~\cite{E143,SMCg2p,E143_2,SMCg2d}, as well as for the 
neutron \cite{E142_2,E154_2}.
In this Letter, we report new measurements of g$_{2}$ and A$_{2}$
for the proton and deuteron made during  experiment E155 at SLAC.

A 38.80 GeV, 120 Hz  electron beam with a longitudinal polarization of 
$(81.3\pm2.0)$\%  struck  transversely polarized
NH$_3$\cite{E143} or $^6$LiD\cite{LiD} targets. The beam helicity direction
was randomly chosen pulse-by-pulse.
Scattered electrons were detected in
three independent spectrometers centered at 2.75$^\circ$, 5.5$^\circ$, and
10.5$^o$.  The two small angle spectrometers
were the same as in SLAC E154 \cite{E154}, while the large angle spectrometer
was new for this experiment. It  consisted of a single dipole magnet
and two quadrupoles, and covered a momentum range from 7 to 20 GeV, and
scattering angle range from $9.6^\circ$ to $12.5^\circ$ with
a maximum solid angle of 1.5 msr at 8 GeV. Electrons
were separated from a much larger flux of pions using a gas \v Cerenkov counter
and a segmented electromagnetic calorimeter. 
Further information on the technique can be found in
references \cite{E143,E154,E155}. 

The measured counting rate asymmetries from the two beam helicities were
corrected for beam polarization, target polarization, tracking efficiencies, 
pion and charge symmetric backgrounds, and radiative effects.
Uncertainties in the
radiative corrections were estimated by varying the input models over
a range consistent with the measured data.
The deuteron  data were extracted from the $^6$LiD results by applying a
correction for both the lithium and deuterium nuclear wave functions with
$^6Li \sim \alpha + d$\cite{LiD}.
 The structure function
g$_{2}(x,Q^{2})$ and the virtual photon
absorption  asymmetry 
A$_2(x,Q^2)$ are usually determined from the two
measurable asymmetries, $A_{\perp}(E,x,Q^2)$ (dominant contribution) and
$A_{\parallel}(E,x,Q^2)$ (small contribution), 
corresponding to transverse  and longitudinal target polarization
with respect to the incoming electron beam helicity. Because in this
experiment these
asymmetries were measured at two different beam energies (38.8 and 48.3 GeV
respectively), we instead chose to
determine g$_2$ and A$_2(x,Q^2)$ from $A_\perp$ (dominant contribution)
and $A_1$ (small contribution) using:
\begin{eqnarray}
{\rm{g}}_2(x,Q^2)&=&\frac{F_2(x,Q^2)}{2x\gamma (1+R(x,Q^2))}
\biggl[A_{\perp}(E,x,Q^2)/d + A_1(x,Q^2)(\zeta -\gamma)\biggr]  \\  
{\rm{A}}_2(x,Q^{2})&=& A_{\perp}(E,x,Q^2)/d  +\zeta A_1(x,Q^2)
%{\rm{g}}_{2}(x,Q^{2})=\frac{F_{2}(x,Q^{2})(1+\gamma ^{2})}
%{2x\left[1+R(x,Q^{2})\right]}
%\frac{y}{2d\,{\rm{sin}}\, \theta}
%\biggl[A_{\perp}\frac{E+E'{\rm{cos}}\, \theta}{E'} 
%-A_{\parallel}(E,x,Q^2)\: {\rm{sin}}\, \theta \biggr] \nonumber \\ \nonumber \\ \nonumber
%{\rm{A}}_{2}(x,Q^{2})=\frac{\gamma (2-y)}{2d\, 
%{\rm{sin}}\,\theta }\left[A_{\perp}
%\frac{y(1+xM/E)}{(1-y)} +  A_{\parallel}\: {\rm{sin}}\,\theta \right].
\end{eqnarray}
where  $\zeta= \eta (1+\epsilon)/(2\epsilon)$, 
$\eta=\epsilon\sqrt{Q^2}/(E-E'\epsilon)$,
$E$  and $E'$ are the incident and scattered electron energies, 
$\gamma=2Mx/\sqrt{Q^2}$, 
$d= (1-E'\epsilon/E)\sqrt{2\epsilon/(1+\epsilon)}/(1+\epsilon R)$, and
%d=(1-\epsilon)(2-y)/y[1+\epsilon R(x,Q^{2})]$, and 
$\epsilon^{-1}= 1+2\left[1+\gamma ^{-2}\right]{\rm{tan}}^{2}(\theta /2)$.  
We used a new $Q^2$-dependent parameterization of $A_1$ using existing 
data ~\cite{SMC,E143}
and data from this experiment\cite{E155}.  
The NMC fit to
$F_{2}(x,Q^{2})$~\cite{NMC} and the new SLAC fit to
$R(x,Q^{2})=\sigma_L/\sigma_T$~\cite{R1998} were used.

Results for A$_2$ and  $xg_2$ for the three spectrometers
are given in Table~\ref{tb:A2-XG2} with statistical  errors. The systematic 
errors were negligible by comparison.
The data cover the kinematic range $0.02\leq x \leq 0.8$ and 
$1.0\leq Q^{2} \leq 30$ (GeV/c)$^{2}$ with an average $Q^2$ of $5$
(GeV/c)$^2$.  Figure ~\ref{fg:q2} shows the  values of $A_2$ as a function of
$Q^{2}$ for several values of $x$ along with results from E143\cite{E143}.
There is no evidence of a $Q^{2}$ dependence  for A$_2$ or
$xg_2$ (not shown) within the
experimental errors so the data from all spectrometers were
averaged.  These averaged results for A$_2$ and  $xg_2$  are shown at
the bottom of  Table~\ref{tb:A2-XG2} and  A$_2$ is presented
in Fig.~\ref{fg:a2} along with the $\sqrt{R}$ positivity limit and data
from previous  experiments.  The data are in good
agreement with the previous measurements and  improve the
accuracy for the deuteron. The combined results are significantly smaller than
the positivity limit  over most of the measured range. $A_2$ is consistent
with $A_2^{WW}$ calculated from $g_2^{WW}$ and  $A_2^p$ is
significantly larger than zero around $x\sim 0.2.$
Results for $x{\rm{g}}_2$ are shown in Fig.~\ref{fg:xg2} along with  
the twist-2 component, $x{\rm{g}}_{2}^{WW}$ calculated 
using our new parameterization of the 
$A_1$ data. The combined SLAC data agrees with  g$_2^{WW}$
with a $\chi^2/$(dof) of \ChiGwwp and \ChiGwwd for proton and deuterium 
respectively for 17 degrees of freedom.
The  comparison with g$_2=0$ has similar agreement with
$\chi^2/$(dof) of \ChiZerop and \ChiZerod respectively.  %   coarse bins
Also shown is the bag model calculation of Stratmann\cite{Stratmann}.
A recent Chiral Soliton Model calculation\cite{WGR} (not shown) also agrees with the
data

We used Eq. \ref{eq:g2bar} to calculate the matrix elements d$_n$ assuming
that  $\overline{g_2}$ is independent of $Q^2$ in the measured region.
This is not unreasonable since d$_n$ is supposed to depend logarithmically
on $Q^2$\cite{Vain}.  The part of the integral for $x$ below the measured
region was assumed to be zero because of the $x^n$ suppression. 
For $x\geq 0.8$ we used $\overline{g_2}\propto (1-x)^m$ where m=2 or 3,
normalized to the data for $x\geq 0.5$. 
Because $\overline{g_2}$ is small at high $x$, the contribution was negligible
for both cases. We obtain values of
$d_{2}^p=$\dtp$\pm$\dtpEr                %-0.004\pm 0.009$ 
and $d_{2}^d=$\dtd$\pm$\dtdEr         %0.005\pm 0.007$ 
at an average $Q^{2}$ of $5$ (GeV/c)$^{2}$.
We combined these results with those from SLAC experiments on the neutron 
(E142\cite{E142_2} and E154\cite{E154_2}) and proton and deuteron 
(E143\cite{E143}) and obtained average values 
$d_2^p=$\dtpAv$\pm$\dtpAvEr     %0.005\pm0.004$ 
and $d_2^n=$\dtnAv$\pm$\dtnAvEr. %   0.000\pm0.011$.  

Figure ~\ref{fg:theory}~ shows the experimental values of $d_2$ for
proton and neutron with their error, plotted along with theoretical 
calculations using Bag Models (Song\cite{Song}, Stratmann\cite{Stratmann},
and Ji\cite{Ji}); QCD sum rules (
Stein\cite{Stein}, BBK\cite{BBK}, Ehrnsperger\cite{Ehrnsperger});
and Lattice QCD\cite{LQCD}.  The results are compatible with all the models 
within the still large errors except for the proton lattice calculation.

We evaluated the Burkhardt-Cottingham integral (Eq. \ref{eq:bc}) 
in the measured region
of $0.02\leq x \leq 0.8$ at $Q^2=5 {\rm (GeV/c)^2}$ by assuming 
that $\overline{g_2}$ is independent of 
$Q^2$ and thus that all the $Q^2$ dependence of $g_2$ is in $g_2^{WW}.$
The results for the proton and deuteron are \BCp$\pm$\BCpEr  %$-0.10\pm 0.05$
and \BCd$\pm$\BCdEr %$0.08\pm0.05$ 
respectively.  Averaging with the E143 results which cover a slightly
more restrictive $x$ range gives \BCAvp$\pm$\BCAvpEr %-0.03\pm0.03$ 
and \BCAvd$\pm$\BCAvdEr.    %      $0.04\pm0.04$
All of these integrals are consistent with the Burkhardt-Cottingham sum rule 
prediction of zero.
However, this does not represent a conclusive test of the sum rule
because the behavior of g$_2$ as $x\rightarrow\!0$ is not known.
We evaluated the ELT integral, {Eq. \ref{eq:elt}, using our data in 
the measured region.
The result at $Q^2=5{\rm (GeV/c)^2}$ is \ELT$\pm$\ELTEr, 
which is consistant with the
expected value of zero. Including the E143 $g_2$ data \cite{E143} improves the
accuracy to  \ELTAv$\pm$\ELTAvEr. Again the extrapolation to $x$=0 is
not known, but in this case the contribution is suppressed by a factor of 
$x.$  

In summary, we have presented a new measurement of A$_2$ and g$_2$ for
the proton and deuteron 
in the kinematic range $0.02 \leq x \leq 0.8$ and $1.0 \leq Q^2
\leq 30$ (GeV/c)$^{2}$.  Our results for $A_2$ are
significantly smaller than the $\sqrt{R}$ positivity limit over most 
of the measured range and data for g$_{2}$ are  consistent
with the twist-2 g$_{2}^{WW}$ prediction.  The values obtained for the
twist-3 matrix element $d_2$ from this measurement and the SLAC 
average are also consistent with zero.  Future measurements at SLAC and 
Jefferson National Laboratory
will significantly reduce the errors and enable us to make more conclusive
statements about the higher twist content of the nucleon.

We wish to thank the personnel of the SLAC accelerator department for
their efforts which resulted in the successful completion of the E155
experiment.  We would also like to thank J. Ralston for 
useful discussions and guidance.  This work was supported by the
Department of Energy; by
the National Science Foundation; by the Kent State University Research
Council (GGP);   and by the Centre
National de la Recherche Scientifique and the Commissariat a l'Energie
Atomique (French groups).\\

\newpage
 %DATA FILE USED:
%/afs/slac/u/ra/ser/e155/g2/bosted_analsum.out

 %SPECTROMETERS USED: MIN-MAX   1   3
 % change \  to  
 %\documentstyle[prl,aps]{revtex}
 %\begin{document}
 \begin{table} 
 \caption{ Results for $A_2$ and $xg_2$  with statistical errors for proton and deuteron 
  at the measured $x$ and $Q^2$ [(GeV/c)$^2$]  for the three spectrometers with E=38.8 GeV. }
\label{tb:A2-XG2}
 \begin{tabular}{rrrrrr}
$x$ & $<Q^2>$ &$A_2^p$ & $xg_2^p$ & $A_2^d$ & $xg_2^d$ \\
 \hline
\multicolumn{6}{c}{$\theta\approx 2.75^\circ$ } \\
\hline
 0.022&  1.15& $ 0.149\pm$ 0.111& $ 0.439\pm$ 0.335& $-0.036\pm$ 0.074& $-0.103\pm$ 0.212 \\
 0.026&  1.32& $-0.020\pm$ 0.032& $-0.069\pm$ 0.088& $ 0.023\pm$ 0.021& $ 0.060\pm$ 0.056 \\
 0.039&  1.56& $-0.034\pm$ 0.025& $-0.090\pm$ 0.057& $ 0.023\pm$ 0.017& $ 0.043\pm$ 0.035 \\
 0.062&  1.94& $ 0.025\pm$ 0.033& $ 0.024\pm$ 0.054& $ 0.012\pm$ 0.021& $ 0.013\pm$ 0.034 \\
 0.099&  2.34& $ 0.016\pm$ 0.046& $-0.020\pm$ 0.054& $ 0.040\pm$ 0.031& $ 0.041\pm$ 0.034 \\
 0.159&  2.71& $ 0.033\pm$ 0.069& $-0.021\pm$ 0.056& $ 0.061\pm$ 0.047& $ 0.024\pm$ 0.035 \\
 0.255&  3.01& $ 0.075\pm$ 0.107& $-0.008\pm$ 0.056& $-0.108\pm$ 0.076& $-0.077\pm$ 0.035 \\
 0.411&  3.25& $-0.004\pm$ 0.212& $-0.049\pm$ 0.056& $ 0.273\pm$ 0.166& $ 0.032\pm$ 0.036 \\
 0.621&  3.37& $-0.842\pm$ 0.594& $-0.123\pm$ 0.054& $ 0.501\pm$ 0.503& $ 0.019\pm$ 0.035 \\
 0.796&  3.42& $-0.294\pm$ 1.734& $-0.019\pm$ 0.042& $-2.329\pm$ 1.460& $-0.043\pm$ 0.024 \\
 \hline
\multicolumn{6}{c}{$\theta\approx 5.5^\circ$  } \\
\hline
 0.072&  3.68& $ 0.185\pm$ 0.177& $ 0.367\pm$ 0.370& $-0.177\pm$ 0.117& $-0.353\pm$ 0.231 \\
 0.104&  4.84& $-0.041\pm$ 0.037& $-0.102\pm$ 0.064& $ 0.034\pm$ 0.024& $ 0.050\pm$ 0.040 \\
 0.161&  6.26& $ 0.019\pm$ 0.032& $-0.022\pm$ 0.042& $ 0.007\pm$ 0.022& $-0.007\pm$ 0.027 \\
 0.256&  7.76& $ 0.098\pm$ 0.044& $ 0.030\pm$ 0.039& $ 0.048\pm$ 0.033& $ 0.018\pm$ 0.025 \\
 0.417&  9.20& $-0.004\pm$ 0.079& $-0.059\pm$ 0.034& $-0.030\pm$ 0.063& $-0.032\pm$ 0.022 \\
 0.625& 10.23& $ 0.021\pm$ 0.211& $-0.027\pm$ 0.026& $-0.011\pm$ 0.186& $-0.015\pm$ 0.017 \\
 0.828& 10.76& $-0.298\pm$ 0.701& $-0.010\pm$ 0.014& $ 1.102\pm$ 0.603& $ 0.012\pm$ 0.008 \\
 \hline
\multicolumn{6}{c}{$\theta\approx 10.5^\circ$ } \\
\hline
 0.168&  9.77& $ 0.034\pm$ 0.071& $ 0.011\pm$ 0.115& $-0.069\pm$ 0.049& $-0.118\pm$ 0.070 \\
 0.258& 13.70& $ 0.049\pm$ 0.061& $-0.001\pm$ 0.070& $ 0.006\pm$ 0.044& $-0.016\pm$ 0.043 \\
 0.432& 20.06& $ 0.100\pm$ 0.082& $ 0.006\pm$ 0.045& $ 0.016\pm$ 0.068& $-0.013\pm$ 0.029 \\
 0.643& 25.46& $ 0.186\pm$ 0.136& $ 0.005\pm$ 0.020& $ 0.357\pm$ 0.133& $ 0.030\pm$ 0.015 \\
 0.841& 29.25& $ 0.670\pm$ 0.309& $ 0.008\pm$ 0.006& $-0.295\pm$ 0.322& $-0.006\pm$ 0.004 \\
 \hline
\multicolumn{6}{c}{AVERAGE                       } \\
\hline
 0.022&  1.15& $ 0.149\pm$ 0.111& $ 0.439\pm$ 0.335& $-0.036\pm$ 0.074& $-0.103\pm$ 0.212 \\
 0.026&  1.32& $-0.020\pm$ 0.032& $-0.069\pm$ 0.088& $ 0.023\pm$ 0.021& $ 0.060\pm$ 0.056 \\
 0.039&  1.56& $-0.034\pm$ 0.025& $-0.090\pm$ 0.057& $ 0.023\pm$ 0.017& $ 0.043\pm$ 0.035 \\
 0.062&  1.99& $ 0.030\pm$ 0.032& $ 0.032\pm$ 0.054& $ 0.006\pm$ 0.021& $ 0.004\pm$ 0.033 \\
 0.101&  3.45& $-0.018\pm$ 0.029& $-0.056\pm$ 0.042& $ 0.037\pm$ 0.019& $ 0.045\pm$ 0.026 \\
 0.161&  5.41& $ 0.026\pm$ 0.027& $-0.019\pm$ 0.033& $ 0.006\pm$ 0.019& $-0.007\pm$ 0.020 \\
 0.256&  7.54& $ 0.084\pm$ 0.034& $ 0.015\pm$ 0.029& $ 0.020\pm$ 0.025& $-0.014\pm$ 0.018 \\
 0.421& 11.75& $ 0.045\pm$ 0.055& $-0.039\pm$ 0.024& $ 0.014\pm$ 0.045& $-0.013\pm$ 0.015 \\
 0.637& 19.62& $ 0.103\pm$ 0.112& $-0.023\pm$ 0.015& $ 0.246\pm$ 0.106& $ 0.009\pm$ 0.010 \\
 0.839& 27.18& $ 0.491\pm$ 0.279& $ 0.002\pm$ 0.005& $-0.071\pm$ 0.279& $-0.004\pm$ 0.004 \\
\end{tabular}
 \end{table}
 %\end{document}

\newpage

\begin{figure}
{\epsfig{figure=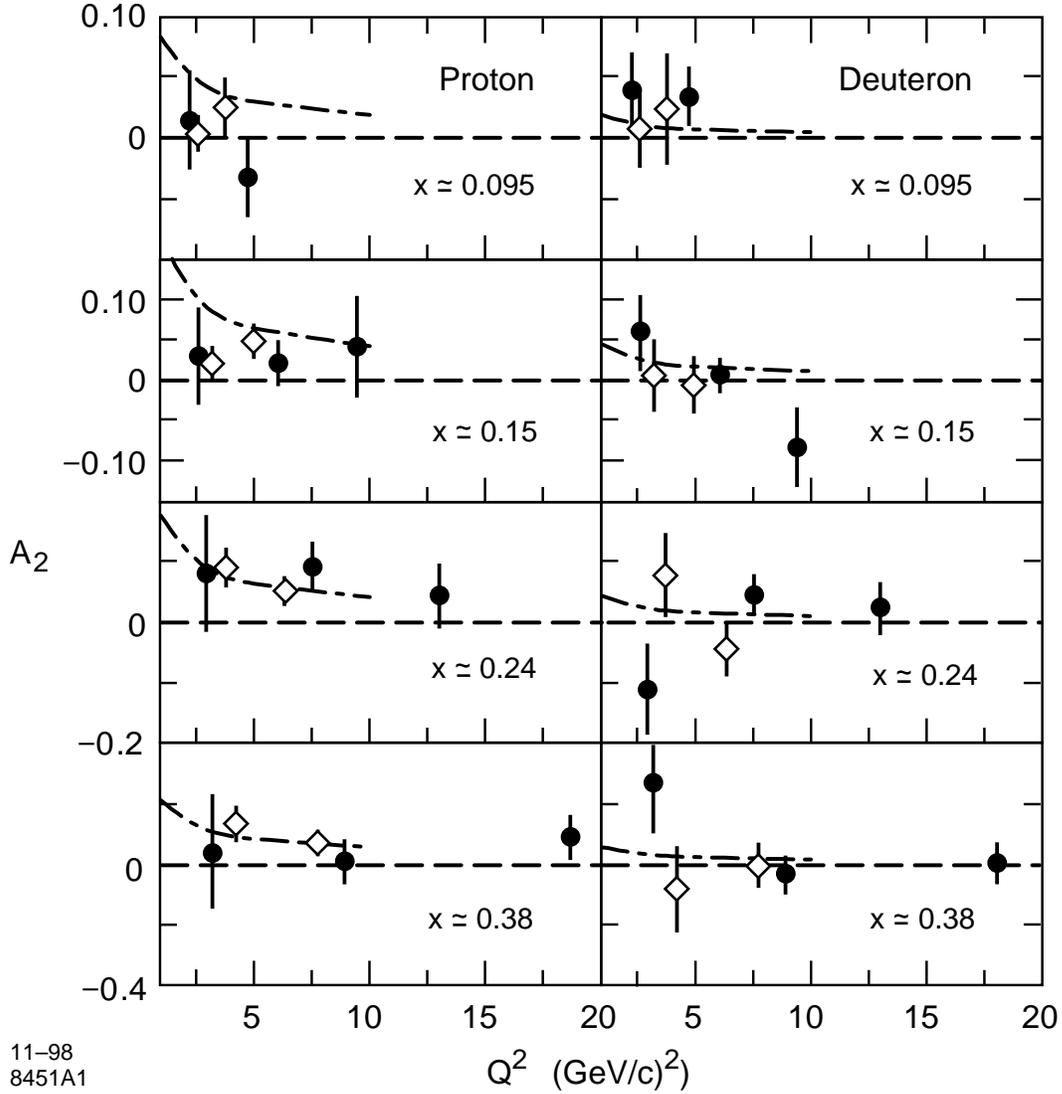,width=14cm}}
%{\epsfig{figure=../AA2-Q2-ALL.ps,width=14.0cm}}
\vskip .2in
\caption{$A_2$ for the proton and deuteron as a function of $Q^2$ for selected
values of $x.$  Data are for this experiment (solid) and 
E143\protect\cite{E143}(open). The errors are statistical; the systematic
errors are negligible. The Bag Model calculation of 
Stratmann\protect\cite{Stratmann} is also shown.  }
\label{fg:q2}
\end{figure}

\begin{figure}
{\epsfig{figure=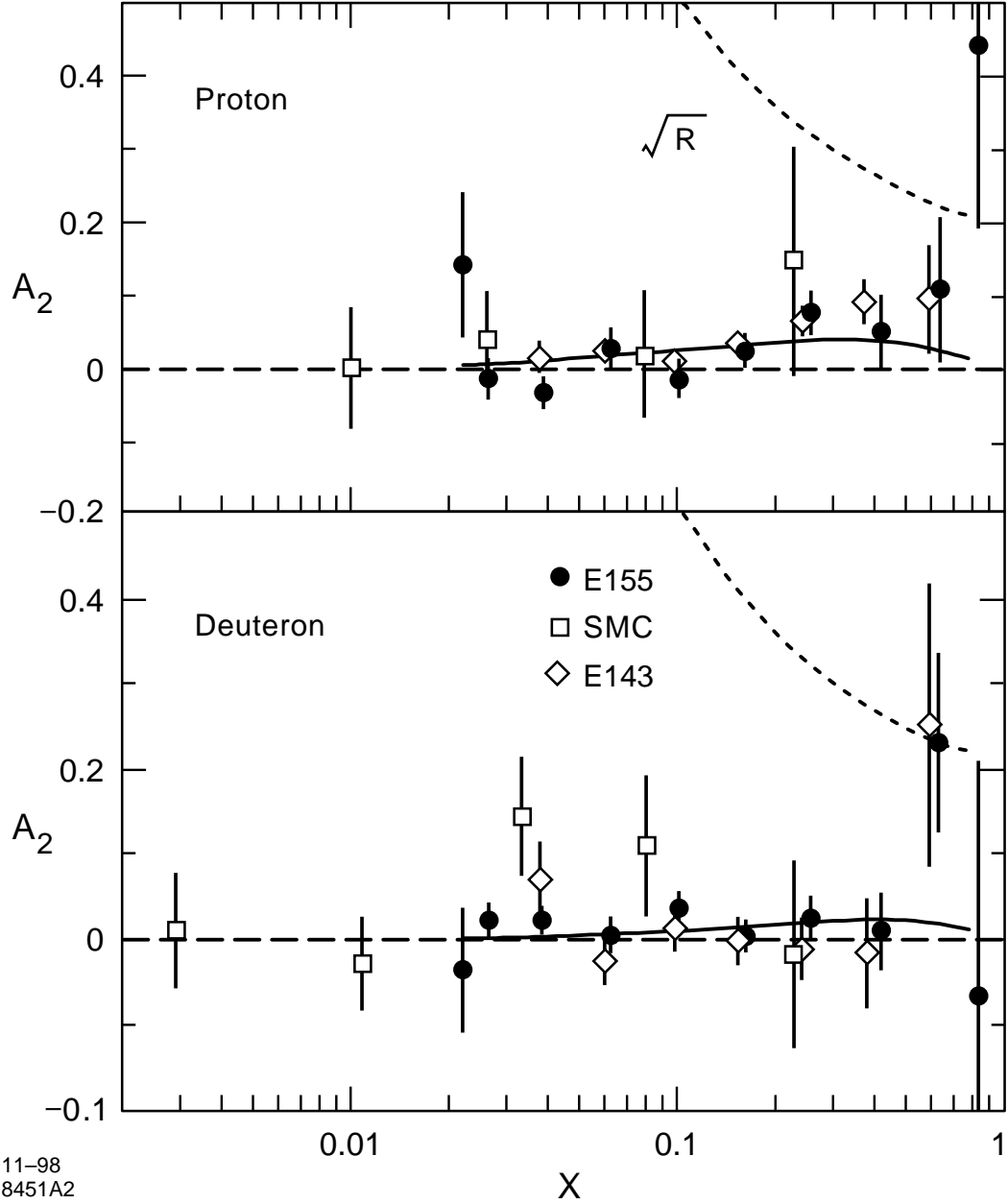,width=14cm}}
%{\epsfig{figure=../A2_SMC.ps,width=14.0cm}}
\vskip .2in
\caption{The asymmetries A$_2$ for proton and deuteron
for this experiment (E155) with data from all spectrometers averaged
(Table \protect\ref{tb:A2-XG2}). The errors are statistical; the systematic errors
are negligible.
Also shown are the  data from SLAC E143 \protect\cite{E143} and 
SMC\protect\cite{SMC}. Our $A_2^{WW}$ calculation is shown as the solid
line and the $\sqrt{R}$ positivity limit is shown as the dotted curve, 
evaluated  at  the average $Q^2$ for this experiment at each $x$. }
\label{fg:a2}
\end{figure}

\begin{figure}
{\epsfig{figure= 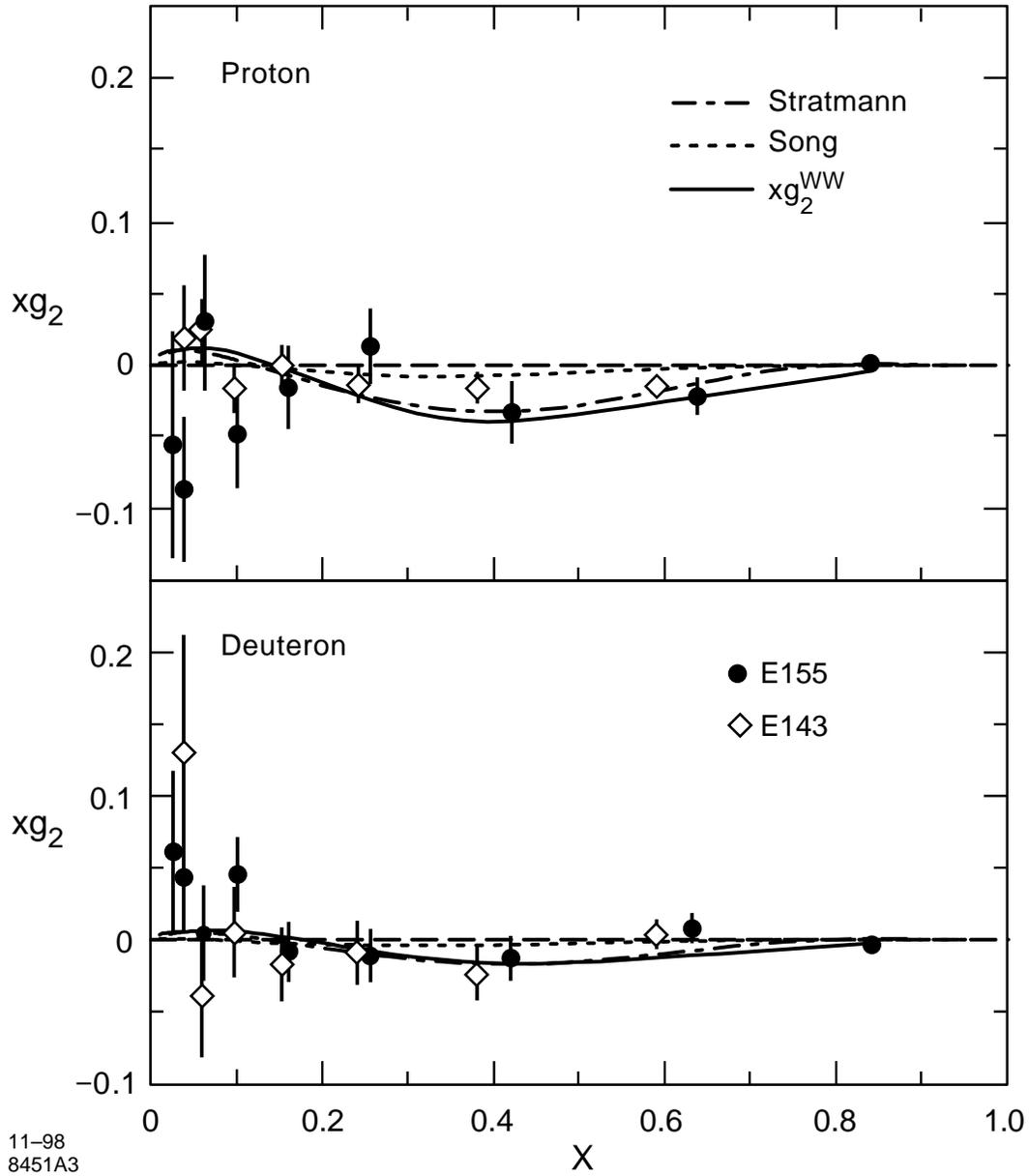,width=14cm}}
%{\epsfig{figure=../xg2w.pswidth=14.0cm}}
\vskip .2in
\caption{The structure function $x$g$_2$ for all spectrometers
combined and data from E143\protect\cite{E143}. The errors are statistical;
the systematic errors are negligible.
Also shown is our twist-2 g$_2^{WW}$ at the average $Q^2$ of this
experiment at each value of $x$ and the calculations of 
Stratmann \protect\cite{Stratmann} and Song \protect\cite{Song}.}
\label{fg:xg2}
\end{figure}

\begin{figure}
{\epsfig{figure=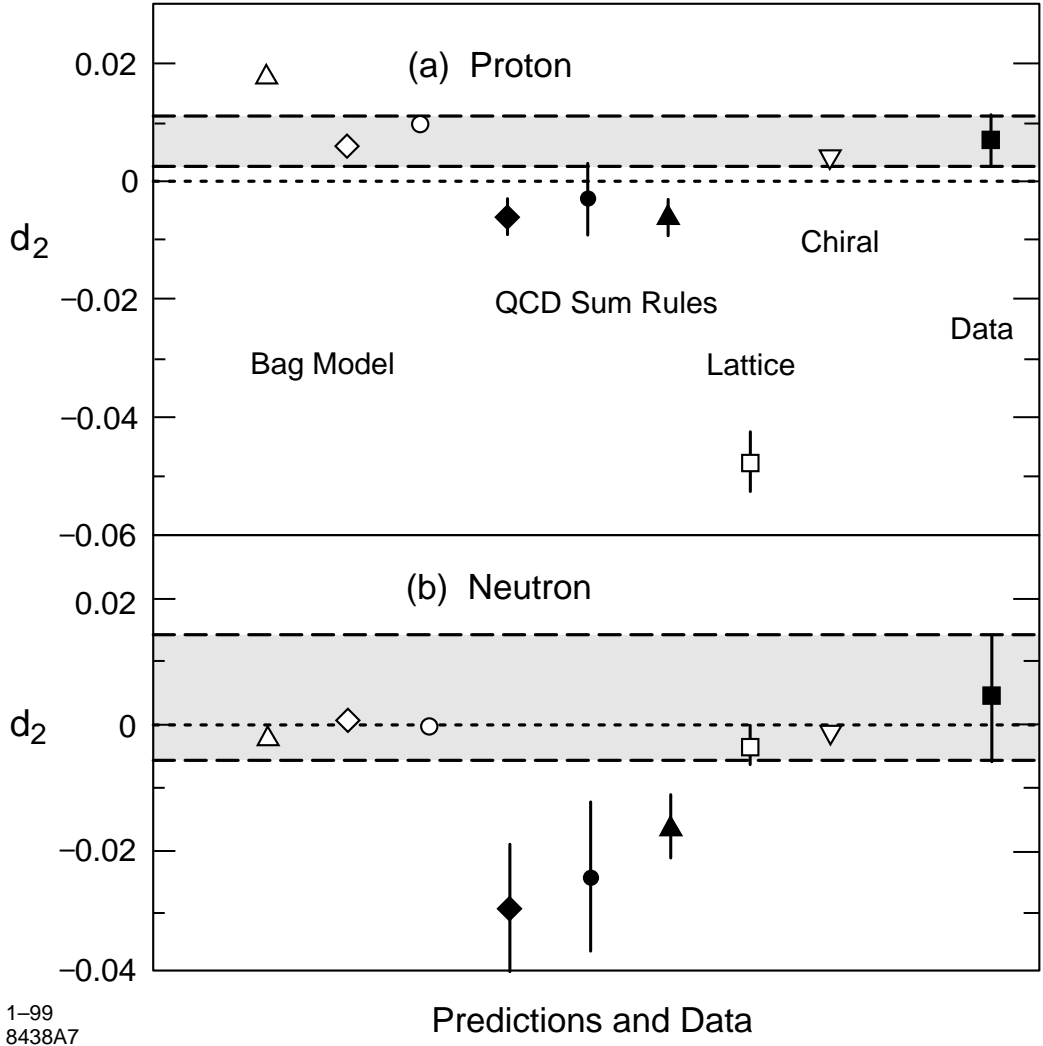}}
%{\epsfig{figure=../theory-plot.ps,width=14.0cm}}
\vskip .2in
\caption{The twist-3 matrix element $d_2$  for the 
proton and deuteron from the combined data from
SLAC experiments E142\protect\cite{E142_2}, E143\protect\cite{E143}, 
E154\protect \cite{E154_2} and E155  (Data).  Also shown are
theoretical models from left to right: 
Bag Models\protect\cite{Song,Stratmann,Ji}, QCD Sum Rules 
\protect\cite{Stein,BBK,Ehrnsperger}, lattice QCD \protect\cite{LQCD}, and
Chiral Soliton Model \protect\cite{WGR}.
The shaded region indicates the experimental errors.}
\label{fg:theory}
\end{figure}

\end{document}